\documentclass[fleqn,usenatbib]{mnras}

\usepackage{newtxtext,newtxmath}

\usepackage[T1]{fontenc}
\usepackage{ae,aecompl}

\usepackage{graphicx}	
\usepackage{amsmath}	
\usepackage{footnote}
\usepackage{longtable}

\usepackage[authoryear]{natbib}
\bibpunct{(}{)}{;}{a}{}{,}
\label{Bibliography}

\def\app#1#2{%
  \mathrel{%
    \setbox0=\hbox{$#1\sim$}%
    \setbox2=\hbox{%
      \rlap{\hbox{$#1\propto$}}%
      \lower1.1\ht0\box0%
    }%
    \raise0.25\ht2\box2%
  }%
}
\def\approxprop{\mathpalette\app\relax}

\title[Energy-dependent emissivity profiles of AGN]{Constraining Energy-dependent emissivity profiles of AGN inflows}

\author[D. I. Ashton et al.]{
Dominic I. Ashton,$^{1}$\thanks{E-mail: d.i.ashton@soton.ac.uk}
\& Matthew J. Middleton
\\
$^{1}$Department of Physics \& Astronomy, University of Southampton, Southampton, SO17 1BJ, UK\\
\\
}

\date{Accepted XXX. Received YYY; in original form ZZZ}

\pubyear{2022}

\begin{document}
\label{firstpage}
\pagerange{\pageref{firstpage}--\pageref{lastpage}}
\maketitle

\begin{abstract}

The emissivity of the accretion flow is a key parameter affecting the shape of both the energy and variability power spectrum of AGN. We explore the energy-dependence of the power spectrum for five AGN, across the {\it XMM-Newton} bandpass, and across the 0.01-1~mHz frequency range, finding a ubiquitous flattening of the power spectrum towards higher energies. We develop a framework to explore this behaviour and thereby extract the energy dependence of the emissivity assuming a simple disc-like geometry for the inflow. We find that the emissivity ranges from $\approxprop~R^{-2}$ at energies around the soft excess and increases to $\approxprop~R^{-4}$ or steeper above 4-6 keV. We describe the changing emissivity index with a linear function in energy, finding the best-fitting slopes to vary between AGN. We attempt to correlate the slope of the linear function against key AGN parameters but, as yet, the sample size is too small to confirm hints of a correlation with Eddington ratio.  

\end{abstract}




\section{Introduction}

Studying the variability of AGN on short ($<$ Ms) timescales provides a means by which to probe the otherwise unresolvable geometry of the inner accretion flow \citep{2009Fabian1HKLlines,Wilkins2013,2019KaraContract, 2020AlstonIRASreverbmap}, search for quasi-periodic oscillations (QPOs) resulting from plasma dynamics or instabilities (\citealt{2008Nature}; \citealt{2010MiddletonBHBAnalogy}; \citealt{2014Alston5}), and explore scaling relations for the SMBH mass (e.g. \citealt{2004PapadakisAGNScaling}; \citealt{2006McHardyAGNScaling}). Where data quality permits, the energy-dependence of the variability can provide even greater insight via, e.g. energy-lag spectra \citep{2019KaraContract}, deeper QPO searches \citep{2021Ashton}, rms and covariance spectra \citep{2009MiddletonSoftExcess, 2011Middleton8Years, 2014UttleyXrayReverb}.   

It has been established that the shape of the power spectrum's broad-band noise is formed as a result of propagation of mass accretion rate fluctuations (\citealt{1997LyubarskiiPropagation}; \citealt{2001Churazov}; \citealt{2011IngramDone}; \citealt{2021ReynoldsPropagation}), and is energy-dependent due to the sampling of different spatial scales, differing emissivities \citep{2006ArevaloUttley}, and the different processes dominating various energy bands (e.g. the soft excess versus bands in which the intrinsic coronal emission dominates). Studies of the energy-dependent shape of the power spectrum have been rather limited, partly due to the requirement for high signal-to-noise data and long observations (to probe down to low frequencies and well-constrain the shape of the power spectrum). When studies have been performed, they are usually restricted to a single AGN of interest, e.g. NGC 7469 (in three energy bands: \citealt{2001NandraNGC7469}), MCG-6-30-15 (in three energy bands: \citealt{2003VaughanMCGvariability, 2005McHardyMCG63015}), NGC 4051 (in two energy bands: \citealt{2004McHardyNGC4051Edepend}, and in three energy bands using concatenated observations totalling $\sim 570$ ks:  \citealt{2011VaughanNGC4051}), and most recently in RE J1034+396 (in three energy bands: \citealt{2020JinREJ}). In all of these previously published cases, the observed trend is that the index of the power-law describing the power spectrum becomes flatter with increasing energy (i.e. there is relatively more variability power at higher frequencies at higher energies). 

In this paper we explore in greater detail, the energy-dependence of the power spectral shape for a sample of AGN. Ignoring reflection, we successfully model the changing power spectral index with energy as a result of energy-dependent emissivities and a low pass filter \citep{2006ArevaloUttley}. The paper is structured as follows: in Section 2 we discuss the AGN sample and the data reduction methods. In Section 3 we describe the modelling of the power spectra. In Section 4 we present our results, showing the energy-dependent nature of these power spectra, across multiple AGN. In Section 5 we present our theoretical framework and in Section 6, we discuss the limitations of our approach and highlight the wider relevance and impact of our results.

\section{Data analysis}

\subsection{AGN Sample and Data reduction}

The data and extraction methods we have used are presented in \citet{2021Ashton} in which we consider a sample of 38 bright AGN and a total of 200 {\it XMM-Newton} observations. The combination of {\it XMM-Newton} with bright AGN results in high signal-to-noise data and typically long (up to 130 ks) exposures. The source sample is drawn from the Palomar Green QSO (PGQSO) catalogue as used by \citet{2006CrummyCatalogue} and \citet{2007MiddletonPGQSRs}, with a number of additional, bright and well-observed AGN: 1H~0707-495, MS~22549-3712, NGC~4151, PHL~1092 and IRAS~13224-3809. The total sample consists of 22 NLS1s and 16 Type-1 Seyferts, as defined by \citet{2010VernonCetty}. 

We considered all observations in the \textit{XMM-Newton} public HEASARC archive\footnotemark\footnotetext{https://heasarc.gsfc.nasa.gov/} up to September 2019, and follow the same data extraction method described in full in \citet{2021Ashton}, which we do not replicate here for the sake of brevity. Of importance is that we choose to use only the PN (rather than PN and MOS), and follow standard procedure by selecting only single/double pixel events (with \textsc{pattern $\le$ 4}), before using the standard \textit{XMM-Newton} data reduction pipeline \footnote{https://www.cosmos.esa.int/web/xmm-newton/sas-thread-timing} in \textit{SAS v17.0.0}. We use \textsc{evselect} to extract source and background light curves (from 40" regions) with a binsize of 100s and a sliding energy window across the nominal $0.3 - 10$ keV range of the instrument. This window is broken into 50 non-overlapping energy bins, allowing for 1225 different energy combinations to be studied (e.g. $0.3 - 0.5$ keV, $0.3 - 0.7$ keV etc.). We then proceed to use {\sc epiclccorr} to subtract the background and apply the necessary corrections to each of the light curves.

Times corresponding to flares in the high-energy ($10 - 12$ keV), full-field background were excised (see \citealt{2021Ashton} for details). Such flare subtraction can introduce gaps into the lightcurves, which, if left unmodified, can affect the shape of the power spectrum. Standard approaches include linear interpolation  (\citealt{2012GonzMartinVaughan}; \citealt{2015AlstonMS2254}) but in our analysis, we take the most conservative approach and consider only the longest continuous segment between flares. Whilst this has the potential to restrict the low frequency end of the power spectrum, we find that many of the longest continuous segments in our sample allow us to reach down to $\sim$0.01 mHz. 

\section{AGN Power Spectra}

\begin{figure*}
    \centering
    \includegraphics[width=0.95\textwidth]{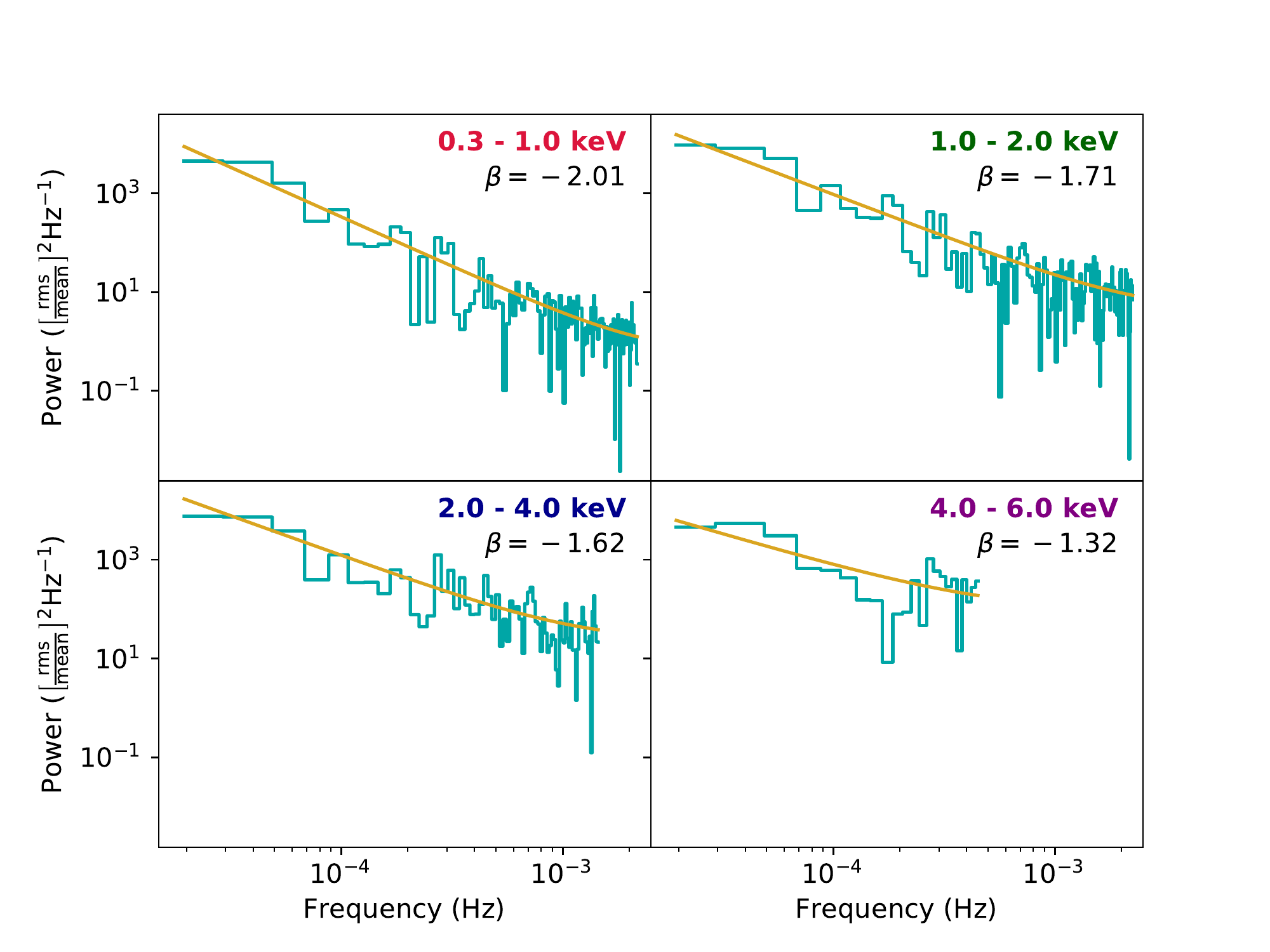}
    \caption{Example energy-resolved periodograms of IRAS 13224-3809 (OBSID:0780561601), fitted using a maximum likelihood \textsc{plc} model, with best-fitting $\beta$ values illustrated. The decreasing number of frequency bins reflects the increasing levels of white noise with increasing energy.}
    \label{fig:PLC_plots}
\end{figure*}

We calculate the periodogram for each energy-resolved lightcurve in fractional-rms units \citep{2003VaughanVariability} -- examples of which are shown in Figure \ref{fig:PLC_plots} for the case of an observation of IRAS 13224-3809. We analyse the periodogram rather than the PSD in order to maximise the number of Fourier frequencies and access the lowest available frequencies in the lightcurve. To ensure sufficient data for modelling, we restrict our analysis to those periodograms containing  $\ge 13$ frequency bins at frequencies below the point where white noise starts to dominate \citep{2021Ashton}.

Similar to \citet{2012GonzMartinVaughan}, 
we fit each periodogram using a maximum likelihood statistic and determine the preferred description by either a power-law + constant ({\sc plc}) or broken power-law + constant model ({\sc bknplc}). The {\sc plc} takes the form of:

\begin{equation}
P(\nu) = N_0 \nu^{\hspace{0.065cm}\beta} + C
\end{equation}

\noindent which consists of free parameters: $\beta$, the spectral index, and $N_0$, the normalisation, as well as $C$, a constant to account for Poisson noise (which we set to the predicted level of the white noise following the method described in \citealt{2003VaughanVariability}). The {\sc bknplc} model is:

\[P_1(\nu) = N_1 \nu^{\hspace{0.065cm}\beta_1} + C \hspace{0.5cm} \textnormal{for } \nu \leq \nu_{\rm b} \]
\vspace{-0.5cm}
\begin{equation}
P_2(\nu) = N_2 \nu^{\hspace{0.065cm}\beta_2} + C \hspace{0.5cm} \textnormal{for } \nu > \nu_{\rm b}
\end{equation}

\noindent At frequencies at, or below the break ($\nu \leq \nu_{\rm b}$), we assume a spectral index $\beta_1$ and normalisation $N_1$, and at frequencies above the break ($\nu > \nu_{\rm b}$), we assume a spectral index $\beta_2$. The normalisation above the break, $N_2$, is calculated directly from the previous parameters. A Bayesian Information Criterion (BIC) test \citep{Liddle2007} is applied to determine the statistically preferred model. As reported in \citet{2021Ashton}, $\sim 10\%$ of the total sample of power spectra which match our data-quality criteria are preferentially modelled with a broken power-law, a proportion in general agreement with \citet{2012GonzMartinVaughan}. Indeed, this 10\% includes the re-identification of break features in observations first reported in \citet{2012GonzMartinVaughan}. To avoid the complicating effects of modelling a break in the periodogram, energy bands where a break is identified are excluded from our subsequent analysis.

In \citet{2021Ashton}, $7$ AGN in our sample of $38$ were found to show statistically significant QPO-like features in their power spectra, when the broadband noise was fitted with a (statistically preferred) \textsc{plc} model. We therefore take the additional precaution of identifying the Fourier bin corresponding to the most prominent outlier in power above the maximum likelihood \textsc{plc} model, removing it, and refitting using a bootstrap-with-replacement method. This approach minimises the bias from an individual, outlier bin on the shape of the power spectrum.

\subsection{Energy-Dependent Power Spectra}

We explore the energy-dependent shape of the best-fitting {\sc plc} model for each AGN in turn. As we lack the necessary data quality to examine the energy dependence of the power spectrum at the highest resolution of our sliding energy window, we obtain our maximum likelihood \texttt{plc} model parameters in five larger energy bins:  $E_{1} = 0.3 - 1.0$ keV, $E_{2} = 1.0 - 2.0$ keV, $E_{3} = 2.0 - 4.0$ keV, $E_{4} = 4.0 - 6.0$ keV and $E_{5} = 6.0 - 10.0$ keV. For each AGN, each observation, and in each energy bin, we obtain the mean of the power spectral index, $\bar{\beta}$ and its 1$\sigma$ error (we are averaging over a sufficient number of observations in each case to provide quasi-Gaussian errors). 

We search across all $38$ AGN in our sample, and report those cases where we have constrained values for $\bar{\beta}$ across a minimum of four energy bins. Due to diminishing data quality at high energies, we are unable to constrain values for $\bar{\beta}$ as we approach 10 keV for the majority of our sample. It is unsurprising that, of the five AGN for which we can constrain $\bar{\beta}$ values across a minimum of four energy bins (IRAS 13224-3809, 1H 0707-495, MRK 766, NGC 4051 and ARK 564), all are bright and amongst the most well-studied AGN by {\it XMM-Newton}. 

\section{Results}

\begin{figure*}
    \centering
    \includegraphics[width=\textwidth, trim={0 1cm 0 2.6cm}]{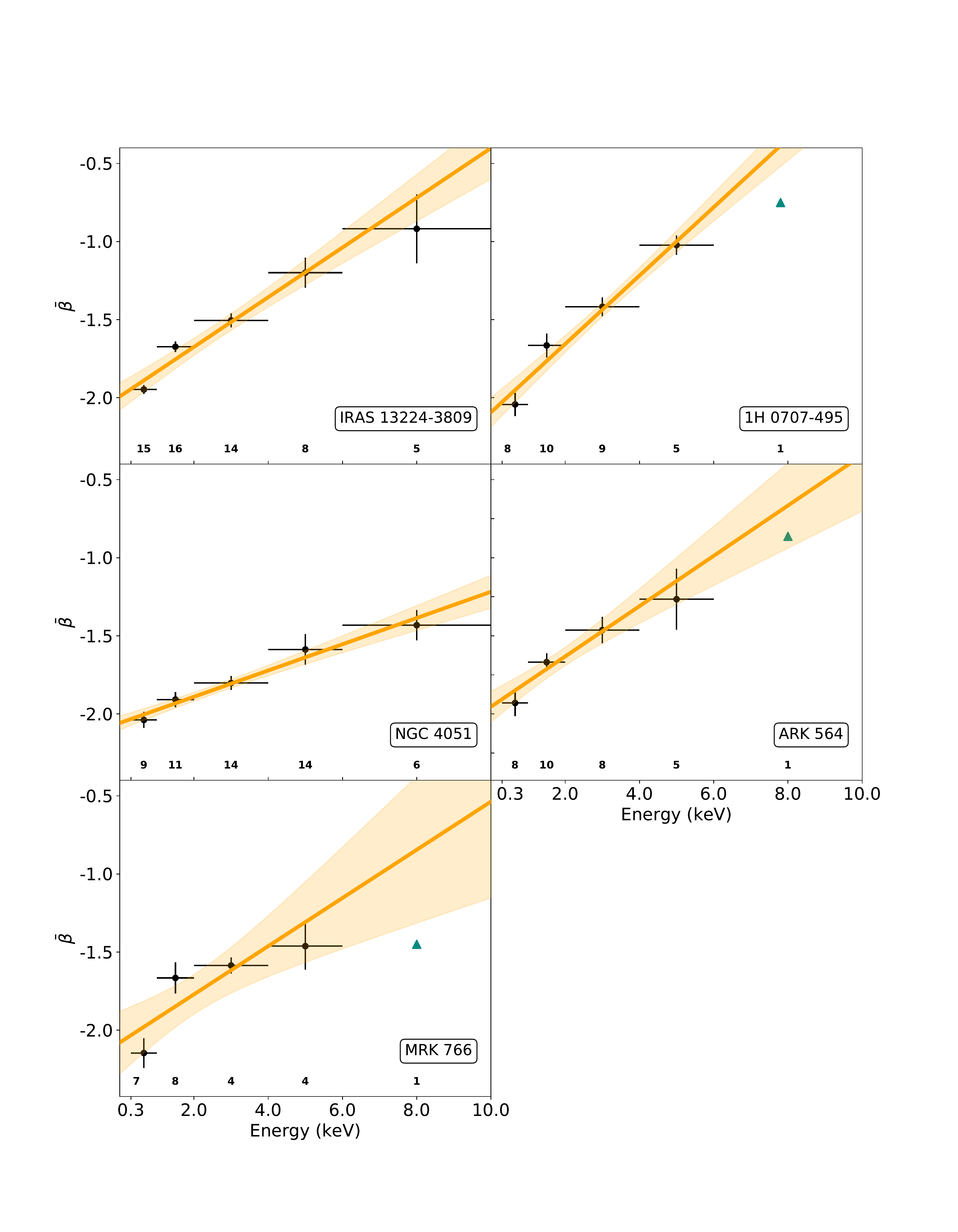}
    \caption{Energy-dependence of the mean power spectral index, $\bar{\beta}$, when modelled with a \textsc{PLC} across the five AGN for which we have constraints in a minimum of four energy bins. We illustrate the maximum likelihood fit and 68\% credible intervals from the MCMC based on a linear model. We show the number of observations that are averaged in order to calculate $\bar{\beta}$ in each energy bin, along the bottom of each panel. In those cases where we are limited to a single observation in an energy bin, we plot the single measurement of $\beta$ (indicated by the filled triangles) and do not use this to contribute to the linear fit due to insufficient statistics.     }
    \label{fig:MCMC_Multiplot}
\end{figure*}

In Figure \ref{fig:MCMC_Multiplot}, we show how $\bar{\beta}$ changes with energy across the five AGN mentioned above. For each AGN, we see that the energy dependence of $\bar{\beta}$ follows an approximately linear trend, with less negative $\bar{\beta}$ values, i.e. a \textit{flattening} of the power spectrum, as we approach higher energies. We proceeded to model the change in index with energy,  $\frac{d\bar{\beta}}{dE}$, for each AGN, with a simple linear model using a widely available Markov-chain Monte Carlo (MCMC) routine (\texttt{emcee}). The high data quality of IRAS 13224-3809 and NGC 4051 allows us to constrain the mean index in the highest energy bin, but for ARK 564, MRK 766 and 1H~0707-495, we are limited to a value from only a single observation in $E_{5}$. We highlight these single points in Figure \ref{fig:MCMC_Multiplot}, but do not use them to constrain our linear model. For the MCMC fitting, we assume uniform priors, and implement a routine with $100$ walkers over 10,000 iterations, with a discarded burn-in time of the first $2000$ steps. We return the median of each parameter's posterior distribution, along with the percentiles required to highlight $68\%$ credible intervals about the MCMC fit, as shown in Figure \ref{fig:MCMC_Multiplot}. The best-fitting $\frac{d\bar{\beta}}{dE}$ values returned by this method are provided in Table \ref{tab:LinearModels}, and the total number of observations contributing to each energy bin for each AGN is indicated in Figure \ref{fig:MCMC_Multiplot}.

As can be seen from Figure \ref{fig:MCMC_Multiplot} and Table \ref{tab:LinearModels}, IRAS 13224-3809, ARK 564 and MRK 766 all have similar values of $\frac{d\bar{\beta}}{dE}$, between $0.15$ keV$^{-1}$ and $0.16$ keV$^{-1}$ which differ compared to NGC 4051 and 1H 0707-495, which have $\frac{d\bar{\beta}}{dE}$ of $0.08~\pm0.01$ keV$^{-1}$ and $0.22 ~\pm0.02$ keV$^{-1}$ respectively. 

Although we have limited confidence in the values for $\bar{\beta}$ above 6 keV for most of our AGN, it is interesting to note that there is a tendency for those values to lie below the linear model, i.e. at steeper indexes than a linear model would predict. It is unclear if this is real but may be confirmed in future using \textit{NuSTAR}. We also note that in all cases, the very softest index lies below the linear relation, indicating that the true description of the energy-dependence likely requires a steeper rise before flattening to higher energies.

\begin{table}
\centering
\caption{Summary of the best fitting $\frac{d\bar{\beta}}{dE}$ values shown in Figure \ref{fig:MCMC_Multiplot}. Columns indicate: (1) AGN, (2) The gradient of the best-fit linear model, with associated 68\% credible interval.}

\begin{tabular}{ccc}
  \hline
  AGN & $\frac{d\bar{\beta}}{dE}$\\
  
 \hline
\vspace{0.12cm}
IRAS 13224-3809 & $0.16~\pm^{0.02}_{0.03}$ ~~ keV$^{-1}$ \\
\vspace{0.12cm}
1H 0707-495 & $0.22~\pm^{0.02}_{0.02}$ ~~ keV$^{-1}$ \\
\vspace{0.12cm}
NGC 4051 & $0.08~\pm^{0.01}_{0.01}$ ~~ keV$^{-1}$ \\
\vspace{0.12cm}
ARK 564 & $0.16~\pm^{0.04}_{0.04}$ ~~ keV$^{-1}$ \\
\vspace{0.12cm}
MRK 766 & $0.15~\pm^{0.04}_{0.05}$ ~~ keV$^{-1}$ \\
 \hline

\label{tab:LinearModels}
\end{tabular}
\end{table}

\subsection{Excluded AGN}

We briefly mention notable AGN where constraints were {\it not} possible for $\geq$ four energy bins. We find that, in many cases, our AGN power spectra can only be constrained at soft energies, typically $< 2$ keV or across a small number of observations in total. An example of this is the case of RE J1034+396, which we find to only have sufficient data to meet our selection criteria below $2$ keV (due to its spectrum being dominated by the soft excess, see \citealt{2001PuchnarewiczREJ1034}; \citealt{2006CasebeerREJ}; \citealt{2009MiddletonSoftExcess}). 

Both PG 1211+143 and RE J1034+396 show the start of a positive linear trend between index and energy (as in Figure \ref{fig:MCMC_Multiplot}) but there are only constraints based on multiple observations in $E_{1}$ and $E_{2}$ for RE J1034+396, and, for PG 1211+143, we have constraints from only a single observation. PKS 0558-504 is similar in appearing to show a weak positive linear trend across three bins but we only have a single observation contributing to the $E_{3}$ bin and none at higher energies for this AGN. 

In the case of PG 0003+199, we observe a strong linear trend across four energy bins ($E_{1} - E_{4}$), comparable to our AGN in Figure \ref{fig:MCMC_Multiplot}, but in this case we are limited to a single observation throughout. Considering only this one observation of PG 0003+199, we would infer $\frac{d\bar{\beta}}{dE} = 0.35$, which is even larger than in the case of 1H 0707-495. However, as with only a single observation we cannot obtain reliable errors for $\frac{d\bar{\beta}}{dE}$, we merely highlight this AGN as worthy of future investigation.

Lastly we report two cases of statistically limited, apparent negative or non-linear trends in PG 1244+026 and PG 1322+659. Both show a flattening in index between $E_{1}$ and $E_{2}$, before a \textit{decrease} in index in $E_{3}$ to values which are even steeper than in the lower energy bins. In both cases however, these trends are derived from only a single observation. 

\section{Theoretical Framework}

The spectra of NLS1s tend to be a complex combination of components (\citealt{2012JinPartIII}; \citealt{2013DoneSpectralComps}), including a soft excess created by partially ionised reflection (\citealt{2006CrummyCatalogue}, \citealt{2007DoneSXSdiscrefl}) and/or Compton upscattering of seed disc photons in a cool corona (\citealt{2009MiddletonSoftExcess}; \citealt{2012DoneIntrinsicDiscEmission}), a power-law from a hot corona and its reflection from the disc, producing -- in addition to a soft excess -- a characteristic Fe K$_{\alpha}$ line and Compton hump (\citealt{1991GeorgeFabianReflection}). Imprinted onto this continuum are lines due to photo-ionised absorption by outflows (\citealt{2010TombesiUFOAGN1}; \citealt{2011TombesiUFOAGN2}; \citealt{2018PintoUFOs}). Each component in the spectrum can vary on a range of timescales and are interlinked either through direct propagation (which produces the canonical rms-flux relationship: \citealt{2001UttleyRmsflux}, and lag to hard energies at low frequencies: \citealt{2013WaltonHardLagsAGN}), or via irradiation (which can yield a soft lag at high frequencies: \citealt{2012ZoghbiSoftLagNGC4151}; \citealt{2016KaraLags}).

Any model which aims to describe the emergent variability must account for the interplay between these components on various timescales. Here, for the sake of simplicity we focus on only the propagation and reflection components, which will be the most salient to our study.

\subsection{Reflection}

Reflection is a well-studied feature of AGN energy spectra (e.g. \citealt{1991GeorgeFabianReflection}) which not only produces the well known Fe line and Compton hump, but can also provide some or all of the soft-excess (e.g. \citealt{2004Fabian1H0707}). Due to being irradiated by an intrinsically variable source of seed photons (the corona), the reflected emission will also vary, with the details of this depending on the energy- and frequency-dependent transfer function (which describes the co-location of the system components in curved space-time).

The potential impact of reflection on the shape of the AGN power spectrum has been investigated by \cite{2016PapadakisPSDs}, \cite{2016EmmanoulopoulosPSDs} and \cite{2019ChainakunPSDs}. Initial investigations focussed on single lamp-post configurations (\citealt{2016PapadakisPSDs}), although no signatures of reverberation could be located in the power spectrum (\citealt{2016EmmanoulopoulosPSDs}). Most recently, \cite{2019ChainakunPSDs} explored how a vertically extended (two-blob) corona can lead to predictable changes due to reverberation, utilising ray tracing in full GR to obtain the response functions. In all cases -- and as expected -- the more extended the corona (in height), the lower the frequency at which the PSD is predicted to change. However, these changes are expected to be small, even where the energy band is dominated by reflection. 

Although the effect of reverberation/reflection on the power spectrum has not been calculated for a radially extended corona (e.g. where the corona is a sandwich type geometry or is the hot innermost flow), the effect can be well reasoned. In all models (regardless of geometry), and for any chosen energy band, $E_A \rightarrow E_B$, the observed lightcurve is a summation $s(E,t) =  p(E,t) + r(E,t)$, where $p(E,t)$ represents emission from the propagation (coronal) component, and $r(E,t)$ from reflection. Assuming that $r(E,t)$ is linearly correlated with $p(E,t)$, we can relate them through an impulse response function $\phi (E,t)$:

\begin{equation}
    r(E,t) = \phi(E,t) \circledast p(E,t)
\label{eqn:impulseresponse}
\end{equation}

\noindent and its Fourier transform (and transfer function $\Phi(\nu)$):

\begin{equation}
    R(E,\nu) = \Phi(E,\nu)P(E,\nu)
\label{eqn:reflection}
\end{equation}

\noindent These components may be expressed in terms of their complex Fourier amplitudes in the standard way:

\begin{align}
&|P|^{2} = P^{*}P \\
&|R|^{2} = R^{*}R = (\Phi P)^{*}(\Phi P) = |\Phi|^{2}|P|^{2}
\label{eqn:Fourier_amplitudes}    
\end{align}

\noindent The final power spectrum we observe is then a weighted sum of these components with distortion due to reflection encoded in the impulse response function, with the lowest frequency effects occurring for the largest light travel times between corona and reflector (regardless of whether the response is an approximation or is ray-traced: \citealt{2019ChainakunPSDs}). In a disc-like coronal geometry (which we will be considering here), the light-travel time between seed photons and reflector is small and so any distortion is similarly small (certainly across the $0.01-1$ mHz frequency range we are investigating here). Given the lack of evidence for reverberation/reflection induced changes to the AGN power spectrum, we will consider only the effects of propagation hereafter.

\subsection{Propagation}

The propagation of material through the accretion flow imprints variability across a broad range of radii (and therefore timescales) by coupling to local mass accretion rate fluctuations (see \citealt{1997LyubarskiiPropagation}; \citealt{2001KotovPropagation}). \citet{2006ArevaloUttley} explored the effect of propagation in acting as a low-pass filter on the shape of the intrinsic power spectrum of the AGN, a process which suppresses high frequency variability. This low-pass filter is described as:

\begin{align}
    \textrm{PSD}_{\rm filtered}(\nu) = \textrm{PSD}_{\rm intrinsic}(\nu) \left(\frac{\int^{r_\nu}_{r_{\rm min}}\epsilon(r) 2\pi r dr}{\int^{\infty}_{r_{\rm min}} \epsilon(r) 2\pi r dr}\right)^2
\label{eqn:propagation_psds}
\end{align}

\noindent where $\epsilon(r)$ is the radial emissivity profile of the flux. From \citet{2006ArevaloUttley}, $\epsilon(r)$ is given by:

\begin{align}
    \epsilon(r) = r^{-\gamma}\left(1 - \sqrt{\frac{r_{\rm min}}{r}}\right)
\end{align}

\noindent with emissivity index $\gamma$, from a minimum radius $r_{{\rm min}}$ out to arbitrary radius $r$. A commonly assumed value for the emissivity index is $\gamma = 3$ (i.e. that of a geometrically thin disc). The value of $\gamma$ may vary broadly with energy, from a suggested minimum of $\gamma \sim 2$ for low energies up to $\gamma > 5$ at the highest energies in the \textit{XMM-Newton} bandpass \citep{2006ArevaloUttley}. Assuming the input power spectrum takes the form of a power-law we can therefore explore the impact of propagation on the changing index of the filtered power spectrum. 

\begin{figure*}
    \centering
    \includegraphics[width=\textwidth]{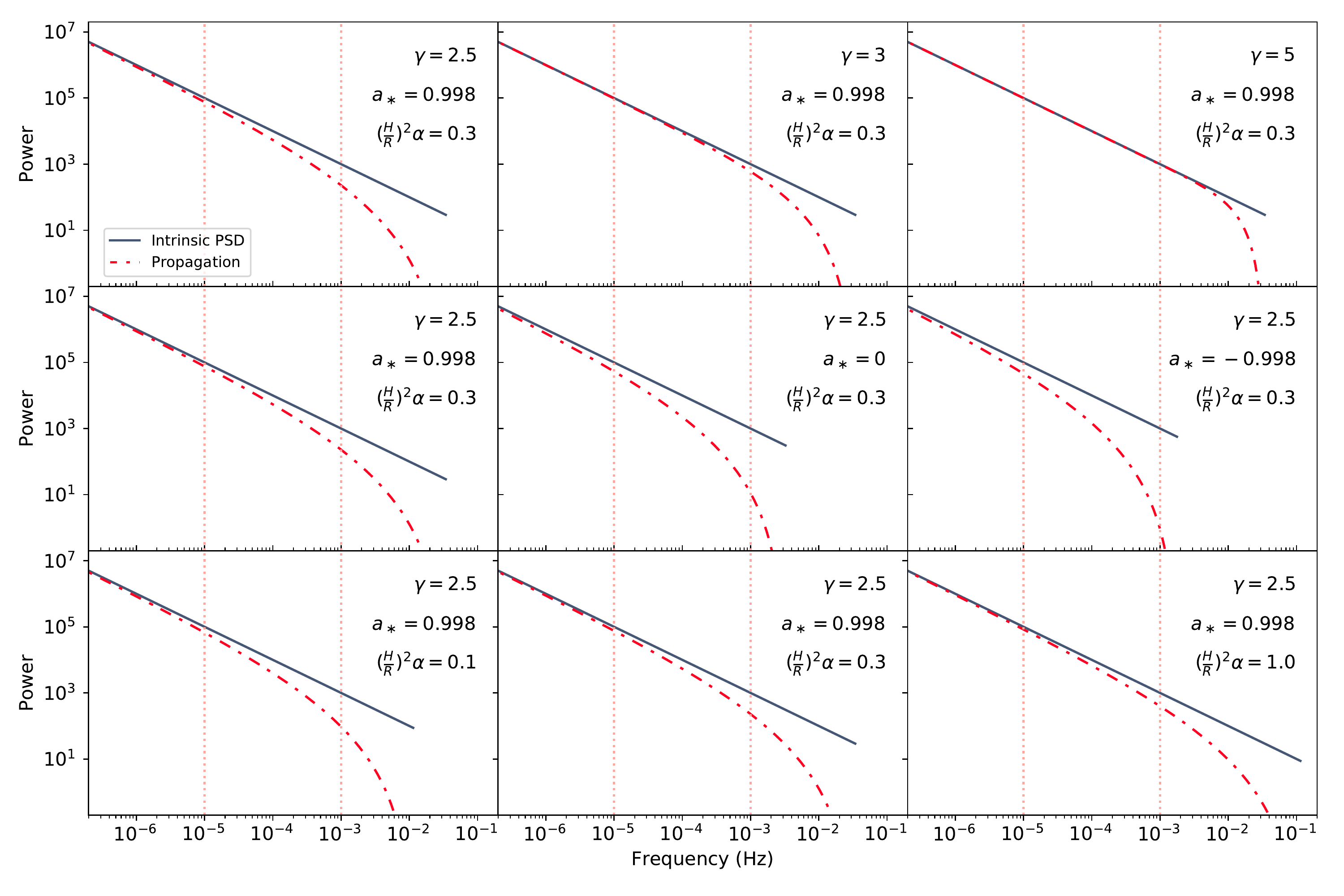}
    \caption{Simulated power spectra for a set of pre-defined emissivity indices, $\gamma$(E), black hole spins, $a_*$ and $\alpha (\frac{H}{R})^2$ values, assuming propagation determines the shape of the broadband noise (distorted from an intrinsic index of $-1$). We show the frequency bands of interest between $0.01-1$ mHz.}
    \label{fig:9paneltheory}
\end{figure*}

\subsection{Modelling propagation}
\label{sec:combinedtheory}

In order to utilise a radial emissivity, as in the model of \citep{2006ArevaloUttley}, we are explicitly assuming a model where the corona is embedded in the disc as a radial structure (see e.g. the model of \citealt{2012DoneIntrinsicDiscEmission}). 

We assume the timescales we observe here are generated in the corona itself (and note that longer timescale changes will be driven by the disc outside of the corona) and are made visible by a changing electron temperature and density as a result of local turbulence and coupling to mass accretion rate propagations through the flow from $r_{\rm out}$ to $r_{\rm in}$. We note that we can discount variations in the seed photon population as the driver for changes in the power spectrum, as the UV emission from the disc is established to be considerably less variable than the corona in NLS1s \citep{1999LeighlyNLS1Variability, 2007SmithUVvariability, 2013AiUVvariabilityNLS1, 2013AlstonUVvariabilityNGC4051, 2016DoneJin1HSpin}. We assume that the variability generated locally at each radius ($r_{\nu}$) is at the viscous frequency (see \citealt{2001Churazov}) such that:

\begin{align}
    r_{\nu} =  \left[ \frac{2\pi\nu}{\alpha}\left(\frac{H}{R}\right)^{-2}\right]^{-2/3} 
\label{eqn:radiusfrequencies}
\end{align}

\noindent where the frequency is in units of $c/R_g$  (see e.g. \citealt{1998KatoAccretionDisks}; \citealt{2006ArevaloUttley}). In the above, $\alpha$ and $\frac{H}{R}$ are the dimensionless viscosity parameter of \citet{1973SS}, and scale height of the accretion disk respectively. We assume that our frequency range of interest, $0.01 - 1$ mHz  (i.e. the range over which which we can practically fit to the data) corresponds to radii between the ISCO ($r_{\rm in}$) and some radius within the true outer edge of the corona (i.e. $r_{\rm out} \le r_{\rm corona}$). The actual frequencies generated in our model therefore depend on the SMBH spin and the combination $(\frac{H}{R})^2 \alpha$ (and somewhat on the SMBH mass -- although here the range is small). Given the reported high spin values for these bright AGN (\citealt{2004NGCLinesSpin};  \citealt{2013FabianIRASspin}; \citealt{2016DoneJin1HSpin}; \citealt{Kara2017ARK564}; \citealt{2018BuissonMRKSpin}), we expect the ISCO to sit at $\sim 1.25 R_{\rm g}$. For the corona at the ISCO to produce variability above our upper frequency limit of 1 mHz requires $(\frac{H}{R})^2 \alpha \gtrsim 0.01$. We note that for the mass range subtended by our AGN sample (from $10^{6.00} - 10^{6.63} M_\odot$, see Table \ref{tab:Gamma_Parameterisation}), should we instead assume zero spin ($r_{\rm in} = 6 R_g$), the viscous frequency at $r_{\rm in}$ is lower and we have strong curvature in our observed bandpass.

As neither $\frac{H}{R}$ nor $\alpha$ are well determined in practice, in the following we simply assume $(\frac{H}{R})^2 \alpha = 0.3$ to be consistent with \citet{2006ArevaloUttley}, and which is appropriate for large scale height inflows (which is a possibility in the inner regions of these high Eddington ratio AGN) and/or large values of $\alpha$. A frequency of 0.01 mHz then corresponds to an outer radius of $\sim 100 R_{\rm g}$ (for a $10^{6} M_{\odot}$ SMBH). It should be recognised that this is sensitive to $\alpha (H/R)^{2}$ and instead, at a value of 0.1, the outer radius is then $\lesssim 50 R_{\rm g}$. We will discuss the possible implications of such large truncation radii in the Discussion.

We assume the intrinsic (unfiltered) power spectrum is well described by a power law with index, $\beta = -1$ which is appropriate for frequencies above any low frequency break (analogous to those detected in black hole XRBs, see e.g. \citealt{2006McHardyAGNScaling}) but below the high frequency break at the ISCO (\citealt{2002PapadakisArk564Break}; \citealt{2006ArevaloUttley}; \citealt{2006McHardyAGNScaling}). Following \citet{2006ArevaloUttley}, and as explained above, we set $r_{\rm in}$ to be the position of the ISCO with $r_{\rm in} = 1.25 R_g$, whilst we set $r_{\rm out} = 100 R_g$ (although we reiterate that this need not be the actual extent of the corona). Following equation \ref{eqn:propagation_psds}, we can obtain the \textit{filtered} power spectrum in a given energy range assuming a value for the emissivity index $\gamma(E)$. To illustrate this process, Figure \ref{fig:9paneltheory} shows a range of simulated, filtered power spectra, all produced from an intrinsic slope of $\beta = -1$, assuming values of $(\frac{H}{R})^2 \alpha$ including  $(\frac{H}{R})^2 \alpha = 0.3$ (following \citealt{2006ArevaloUttley}), and using various parameter combinations for $\gamma(E)$ and $a_*$. In practice, we limit the output (filtered) power spectrum to a frequency range of $0.01 - 1$ mHz, corresponding to the range of frequencies typically accessible in our observations and fit a power-law to each simulated power spectrum and obtain a new value of the power-law index, $\beta$ (corresponding to the effect of propagation).

As can be seen from Figure \ref{fig:9paneltheory}, we are able to obtain substantial energy-dependent changes in $\beta$, obtaining less negative (flatter) values at higher energies, similar to observation (Figure \ref{fig:MCMC_Multiplot}). For a given $\gamma(E)$, we can therefore produce expectation values for $\bar{\beta}$, or conversely use our observed $\bar{\beta}$ to constrain $\gamma(E)$ in this simplified picture. As the exact profile of $\gamma(E)$ is uncertain, we parameterise it in both a linear form:

\begin{equation}
    \gamma(E) = \frac{d\gamma}{dE} E + b
\label{eqn:linearparam}
\end{equation}

\noindent and power law form:

\begin{equation}
    \gamma(E) = \gamma_0 E^{\alpha}
\end{equation}

\noindent and determine the values of parameters $(\frac{d\gamma}{dE}, b)$ and $(\gamma_0, \alpha)$ which maximise the likelihood function between our observed values of $\bar{\beta}_{\rm obs}$ and those theoretical values extracted via the above process $\bar{\beta}_{\rm mod}$. This is equivalent to minimising the negative log-likelihood, $S \equiv -2 \ln{[\mathcal{L}]}$ \citep{2005Vaughan}. Assuming emissivity indices $\hat{\gamma}(E_{\rm k})$ across energy bands k$ = 1,2,...,n-1$, $S$ may be written as:

\begin{equation}
S = 2 \sum_{k=1}^{n-1} \left\{ \ln[\bar{\beta}_{\rm mod, \rm k}] +\frac{\bar{\beta}_{\rm obs, \rm k}}{\bar{\beta}_{\rm mod, \rm k}} \right\}
\end{equation}

\begin{figure*}
    \centering
    \includegraphics[width=0.98\textwidth, trim={0 0cm 0 0cm}]{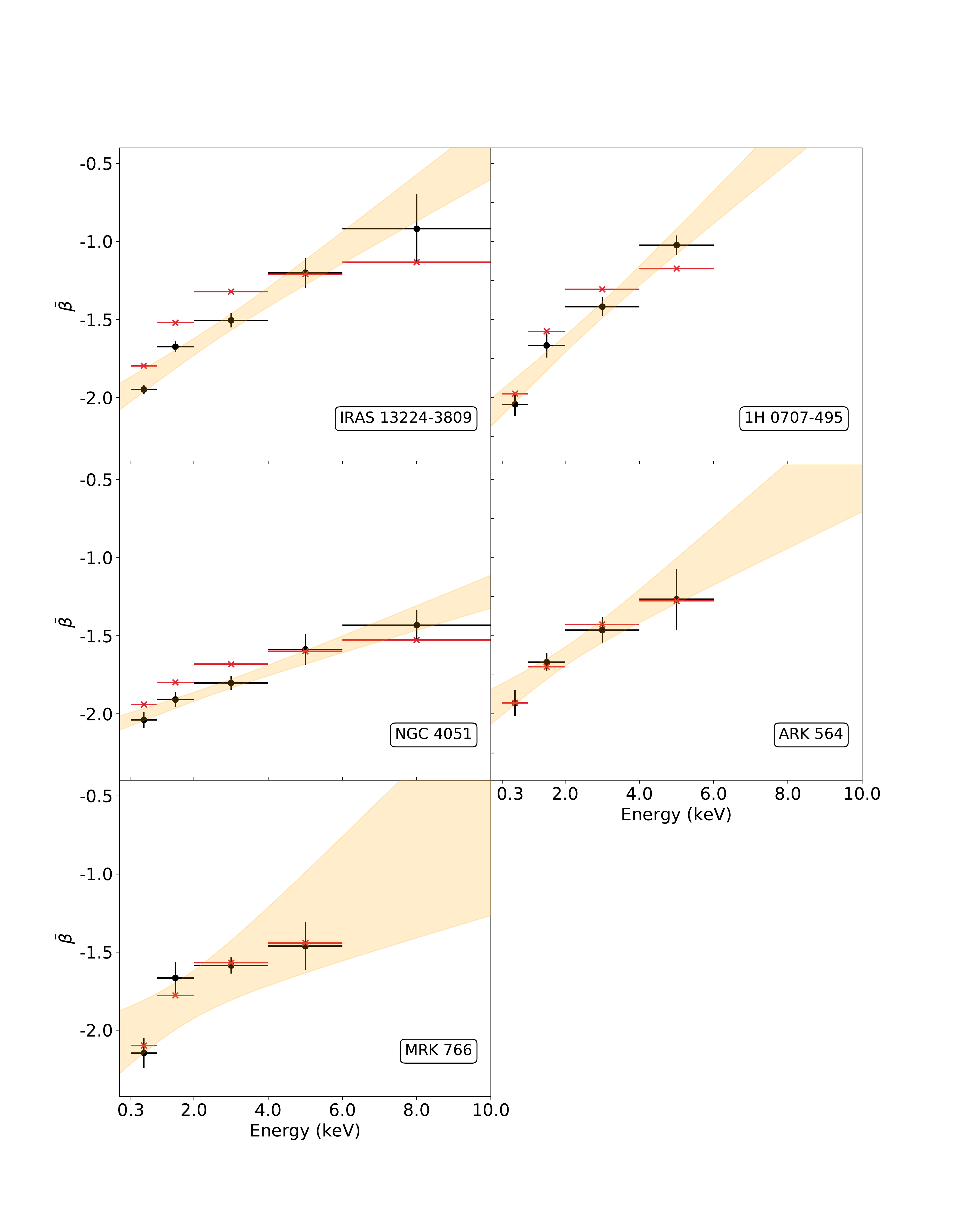}
    \caption{A comparison of mean power spectral index values, $\bar{\beta}$, as a function of energy between our model (red) and observed values (black), across the five AGN. $68\%$ credible intervals are shown as in Figure \ref{fig:MCMC_Multiplot}. We determine the best-fitting linear model, $\gamma(E) = \frac{d\gamma}{dE}E + b$, and select values for $\gamma(E)$ that maximise the likelihood by comparing the theoretical to observed values of $\bar{\beta}$.}
    \label{fig:indexenergytheory}
\end{figure*}

We obtain similarly satisfactory values of the negative log likelihood for both models, with a breakdown of our results provided in Table \ref{tab:Gamma_Parameterisation}. Given the small statistical improvement afforded by the linear model (the fit statistic $S_{\rm lin} \lesssim S_{\rm PL}$ in most cases), we proceed to use this model for $\gamma(E)$ hereafter. We plot a direct comparison between the data and simulated results for the linear model in Figure \ref{fig:indexenergytheory}. 

\begin{table*}
\centering
\caption{AGN properties and results from fitting for $\gamma(E)$, with maximum likelihood determined parameter values $(\frac{d\gamma}{dE},b)$ and $(\gamma_0,\alpha)$, and associated negative likelihood values $S_{\rm lin}$ and $S_{\rm PL}$, for linear and power law models respectively, for each AGN in our study meeting our data criteria. Black hole masses, $M_{\rm BH}$ and bolometric luminosities, $L_{\rm bol}/L_{\rm Edd}$ (calculated using the mass estimate), as well as spin, $a_*$, and inclination angle $i$, are sourced from the literature: (a) \citet{2019AlstonIRASvariability}, (b) \citet{2018BuissonIRAS}; (c) \citet{2013FabianIRASspin}; (d) \citet{2016DoneJin1HSpin}; (e) \citet{2012Dauser1Hxspec}; (f) \citet{Denney2010}, (g) \citet{2002WooUrryBolLum}; (h) \citet{2004NGCLinesSpin}; (i) \citet{2006ZhangARK}; (j) \citet{2007VasudevanBolLum}; (k) \citet{Kara2017ARK564}; (l) \citet{Wang2001}; (m) \citet{2010VasudevanBolLum2}; (n) \citet{2018BuissonMRKSpin}.}

\begin{tabular}{ccccccccccc}
  \hline
  AGN & $\log_{10}(\frac{{}{M_{\rm BH}}}{M_\odot})$ & 
  $L_{\rm bol}/L_{\rm Edd}$  & $a_*$ & $i (\rm deg)$ & $\frac{d\gamma}{dE}$ & $b$ & $S_{\rm lin}$ & $\alpha$ & $\gamma_0$ & $S_{\rm PL}$ \\
  
 \hline
\vspace{0.13cm}
IRAS 13224-3809 & $6.0$ \ (a) & 2.96 
\ (b) & $0.989 \pm 0.001$ \ (c) & $62.3 \pm 1.3$ \ (c) &   0.29 & 1.83 & 13.41 &      0.22 & 2.23 & 13.45 \\
\vspace{0.13cm}
1H 0707-495 & $6.3$ \ (d) & 1.18 
\ (d) & $0.998^{+0.000}_{-0.002}$ \ (e) & $52.0^{+1.7}_{-1.8}$ \ (e) &     0.53 & 1.73 & 11.21 &        0.32 & 2.37 & 11.22 \\
\vspace{0.13cm}
NGC 4051 & $6.2$ \ (f) & 0.19 
\ (g) & $0.973^{+0.025}_{-0.000}$ \ (h) & $48.0^{+4.0}_{-3.0}$ \ (h) &       0.11 & 1.99 & 15.54 &      0.11 & 2.16 & 15.55 \\
\vspace{0.13cm}
ARK 564 & $6.4$ \ (i) & 0.59 
\ (j) & $0.998^{+0.000}_{-0.000}$ \ (k) & $22.0^{+7.0}_{-2.0}$ \ (k) &       0.35 & 2.11 & 11.57 &      0.29 & 2.36 & 11.57 \\
\vspace{0.13cm}
MRK 766 & $6.6$ \ (l) & 0.59 
\ (m) & $0.92^{+0.078}_{-0.000}$ \ (n) & $46.0^{+1.0}_{-2.0}$ \ (n) &       0.36 & 2.31 & 12.25 &       0.22 & 2.75 & 12.24 \\
 \hline

\label{tab:Gamma_Parameterisation}
\end{tabular}
\end{table*}

\begin{figure*}
    \centering
    \includegraphics[width=\textwidth, trim={0 1cm 0 2.4cm}]{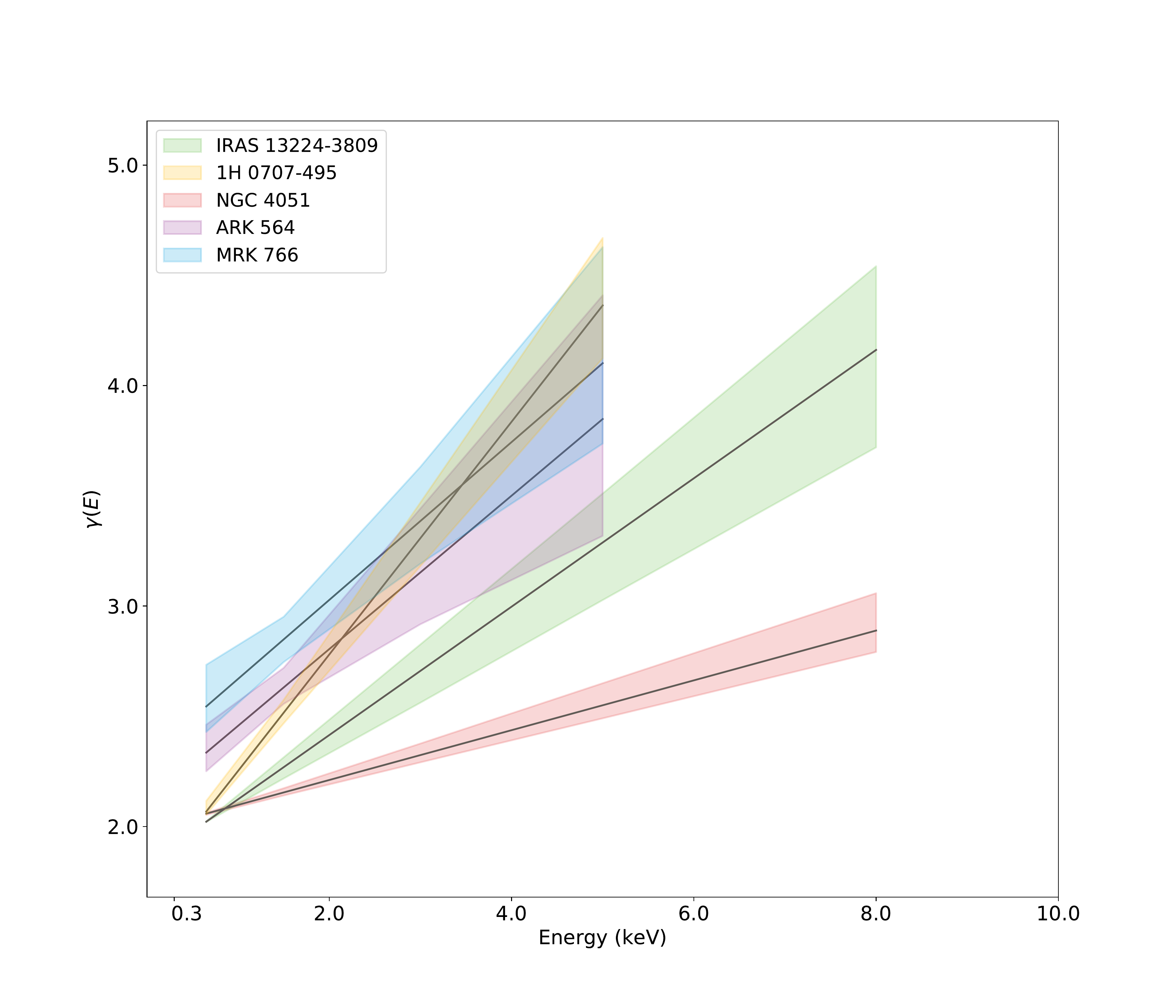}
    \caption{The maximum-likelihood determined values for the emissivity index function, $\gamma(E)$, calculated from the $\bar{\beta}$ values observed across each AGN (as in Figure \ref{fig:indexenergytheory}), each with simulated $\sim68\% \ (1\sigma)$ confidence intervals. All calculations assume a linear parameterisation, $\gamma(E) = \frac{d\gamma}{dE}E + b$.
    }
    \label{fig:emi_energy}
\end{figure*}

As can be seen from Figure \ref{fig:indexenergytheory}, our simple linear model for $\gamma(E)$ provides a reasonable description for the changes in power-law index across all five AGN, and performs extremely well in the case of ARK 564. The model is not capable of providing an ideal description in each case however, with clear systematic offsets visible in the cases of IRAS 13224-3809, 1H 0707-495 and NGC 4051.

The differing nature of the best-fitting emissivity profiles for each AGN is visible from Figure \ref{fig:emi_energy}. The uncertainty on the $\gamma(E)$ values is evaluated through a Monte-Carlo routine, whereby sets of $\beta$ values (equal in number to the number of observations per energy band), are drawn for each AGN across all energy bands (see Figure \ref{fig:MCMC_Multiplot}) which yields an individual MC value for $\bar{\beta}(E)$, with this routine assuming Gaussian errors on the observed values of $\bar{\beta}(E)$, exactly as in Figure \ref{fig:MCMC_Multiplot}. This process is repeated $1000$ times and for each set of $\bar{\beta}(E)$ values, for each AGN, a maximum likelihood set of $\gamma(E)$ values is determined which best describe it. Figure \ref{fig:emi_energy} shows both the maximum likelihood set of $\gamma(E)$ values determined from the observed power spectra and these simulated $68\%$ confidence intervals, where notably, the emissivity appears considerably flatter with energy in NGC 4051 relative to IRAS 13224-3809 and 1H 0707-495 in particular. We speculate as to the origin of such differences in the Discussion.

\section{Discussion}

Modelling our sample of AGN indicates that, in all cases where the data allow us to explore the changing shape of the power spectrum, there is a positive correlation between power spectral index between $0.01 - 1$ mHz (where the periodogram can be well described as a single power-law) and energy (across the $0.3 - 10$ keV range), such that the index flattens with increasing energy. Assuming a simple model of propagating fluctuations -- which is observationally supported by the presence of hard lags at low frequencies in AGN \citep{2013WaltonHardLagsAGN}, the ubiquitous linear rms-flux relationship \citep{2001UttleyRmsflux}, and log-normal flux distribution \citep{2004GaskellLogNorm} -- we have been able to obtain corresponding constraints on the emissivity as a function of energy, (discounting reflection -- see \citealt{2016EmmanoulopoulosPSDs}).

Our model of an in-flow corona (e.g. \citealt{2012DoneIntrinsicDiscEmission}) allows us to assume that the inflow acts a low pass filter, weighted by the radial emissivity profile. We have found that our model requires the emissivity to be around $R^{-2}$ for energies around the soft excess before increasing to a maximum of $\sim R^{-5}$. It is perhaps intriguing to note that radiation pressure or advection dominated flows are expected to have flatter emissivity profiles (with $\gamma < 3$: e.g. \citealt{2000WataraiSMBHs}) which may be consistent with a high accretion rate and such a component producing some of the soft excess (\citealt{2009MiddletonSoftExcess}; \citealt{2012DoneIntrinsicDiscEmission}). Alternatively, it may be that the corona has a larger scale-height (perhaps due to radiation pressure), which pushes the variability to higher frequencies (as the viscous timescale goes as $(\frac{H}{R})^{-2}$), flattening the index to a fixed value above a certain energy. Indeed, we can see from Figure \ref{fig:MCMC_Multiplot} that the index appears to saturate and possibly begin to inflect; whilst we presently lack the statistics to constrain this inflection, with future observations using instruments with higher energy coverage e.g. \textit{NuSTAR}, we will be able to investigate this further. Finally, whilst our analysis assumes an input power-law index of $-1$, other values for the shape of the intrinsic noise are excluded by virtue of producing incompatible results (section \ref{sec:combinedtheory}).

We find our best-fitting function for $\gamma(E)$ to vary across our AGN sample (Figure \ref{fig:emi_energy}). It is plausible that such differences could be due to geometrical effects (e.g. different scale-heights) or some bulk property of the AGN, notably accretion rate, mass and spin. We obtain the spins, inclinations, black hole masses and Eddington scaled accretion rates using $L_{\rm bol}$ from the literature -- see Table \ref{tab:Gamma_Parameterisation}. The spin and inclination values are derived from use of reflection models which assume a lamp-post geometry and as such are not consistent with our assumed geometry. By plotting these spin and inclination values against our derived emissivity gradients we are instead exploring whether an underlying degeneracy might exist. In addition, statistical errors on each parameter are very likely to be underestimates (e.g. see \citealt{2014ReynoldsBHSpin,2016MiddletonAGNspin}); these values are therefore used with due caution.

\begin{figure*}
    \centering
    \includegraphics[width=\textwidth]{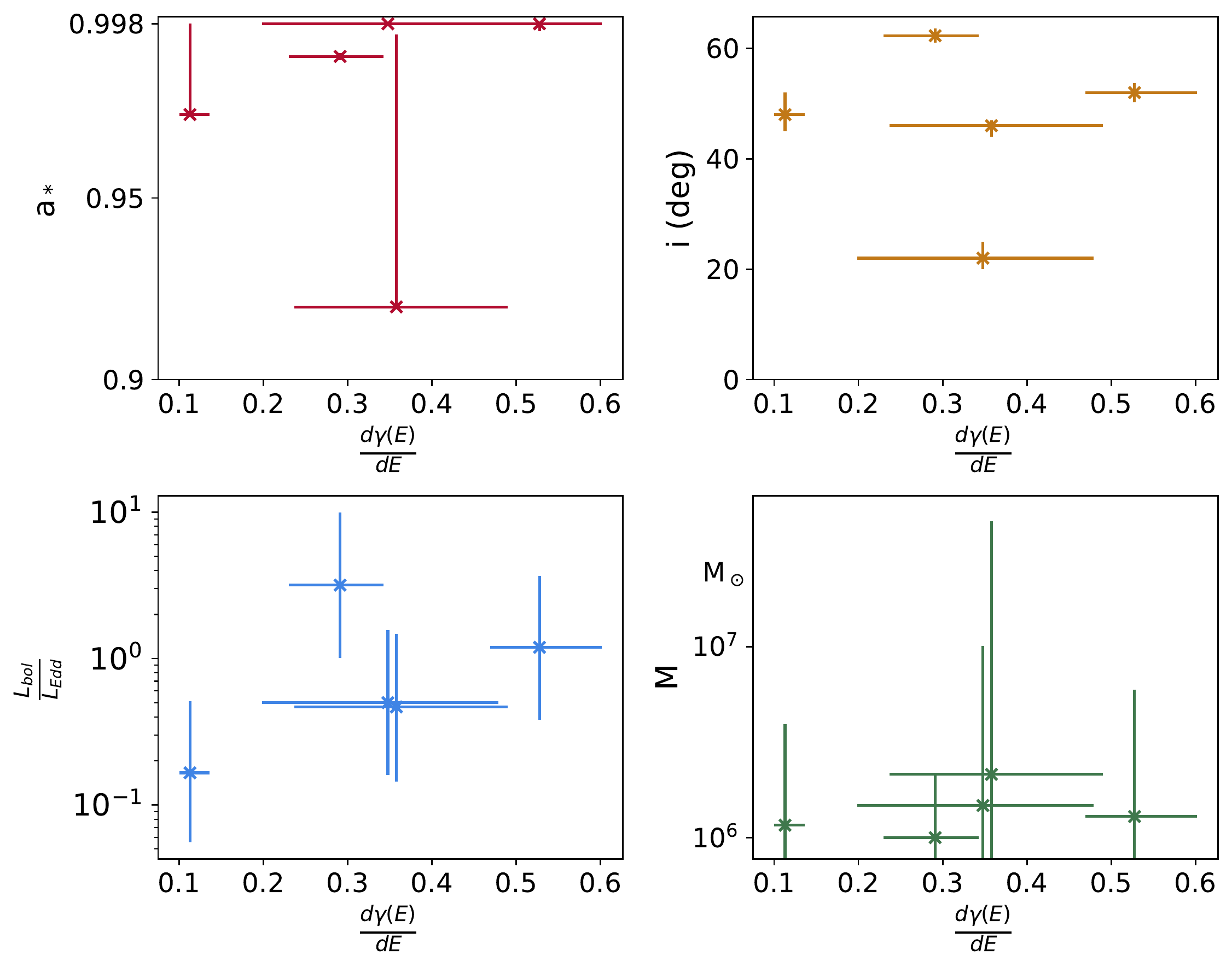}
    \caption{Key macro-properties for our AGN sample -- spin, inclination, Eddington ratio and BH mass plotted against $\frac{d\gamma}{dE}$ (see Table \ref{tab:Gamma_Parameterisation}).}
    \label{fig:macro_properties}
\end{figure*}

In Figure \ref{fig:macro_properties} we plot  $\frac{d\gamma}{dE}$ versus the AGN parameters of interest and find a lack of a correlation with spin, inclination or BH mass but possible hints of a positive correlation with Eddington ratio. In examining the Eddington ratio, there appears to be slightly stronger evidence of a correlation, with the only outlier being IRAS 13224-3809 with a highly uncertain value for $L_{\rm bol}/L_{\rm Edd}$ (see \citealt{2018BuissonIRAS}). We note that the Eddington ratio is also highly uncertain for 1H 0707-495 \citep{2016DoneJin1HSpin}; clearly both a considerably larger AGN sample is required as well as a more accurate handle on the key macro-properties before any claim of a correlation can be made.

One of the most important assumptions in our analysis is the geometry which dictates how our variability is filtered and modified by an assumed radial emissivity (c.f. \citealt{2006ArevaloUttley}). Our assumption here of a continuous disc-like inflow is in contrast to models where the corona sits above the black hole and is probably the base of a jet \citep{2015WilkinsGallocoronajet}. Such lamp-post geometries lead to increased emissivities in the inner portions of the disc -- and therefore the reflected emission -- due to light-bending (\citealt{2011Wilkins1Hemissivity}; \citealt{2017GonzalezEmissivityCoronae}). Caution is needed in assuming a disc-like coronal geometry; whilst an extent of $\sim$ 50R$_{\rm g}$ would be broadly consistent with micro-lensing studies \citep{2013MosqueraLensing}, the magnitude of hard lags may imply considerably smaller coronae. In addition, a radially extended corona of the type we have utilised here would clearly be at odds with relativistic broadening of Fe lines (\citealt{2009Fabian1HKLlines}) unless for some reason the disc reappeared at radii closer to the ISCO.

Importantly, the variability of the corona (regardless of geometry) is subject to other heating/cooling effects besides propagation which are beyond the scope of this paper. Such a model may well explain the changing shape of the power spectrum with energy {\it without} the need to invoke a changing emissivity. Finally, in future it will be interesting to compare the changing shape of the power-spectrum with energy in our AGN sample, to that seen in XRBs (in particular the very high state).

\section{Conclusion}

The well-evidenced propagation model for accretion (\citealt{1997LyubarskiiPropagation}; \citealt{2001KotovPropagation}; \citealt{2011IngramDone}; \citealt{2021ReynoldsPropagation}) makes a clear prediction that -- where we assume a simple disc-like geometry for the corona -- the shape of the power spectrum we observe should be modified by the area of the inflow producing the variability and the radial emissivity profile as a function of energy \citep{2006ArevaloUttley}. Modelling the energy-dependence of the power spectrum should therefore provide interesting insights into the changing radial emissivity under the conditions of the assumed geometry.

Following from the analysis of $38$ bright AGN in \citet{2021Ashton}, we model the energy-resolved periodograms using {\it XMM-Newton} data across a frequency range of 0.01-1 mHz (at frequencies typically lower than the viscous timescale at the ISCO for the high spins and low masses of the AGN in our sample). We apply strict data criteria (both in rejecting observations with an insufficient number of Fourier bins at frequencies below the white noise cutoff, and sources for which there were too few observations to provide error estimates), leaving us with five AGN: IRAS~13224-3809, 1H 0707-495, NGC 4051, ARK 564 and MRK 766. In all five cases, we observe that the power-law index -- which well describes the power spectrum in this frequency range (and is preferred over a broken power-law according to maximum likelihood fitting) -- is steep at low energies and becomes increasingly flat towards higher energies. To-date this has only been reported for a very small number of AGN (with only one such AGN within our sample -- NGC 4051, see \citealt{2004McHardyNGC4051Edepend}; \citealt{2011VaughanNGC4051}) and across only a small number of energy bins \citep{2001NandraNGC7469, 2003VaughanMCGvariability, 2005McHardyMCG63015, 2020JinREJ}; our analysis therefore represents a much broader investigation into the energy-dependence of the shape of AGN power spectra.

Adopting the framework of \citet{2006ArevaloUttley}, we model the power-spectra of our AGN sample in five energy bands, assuming energy-dependent changes are driven by propagation only, and extract the emissivity index as a function of energy, parameterising it as a linear function. This modelling is remarkably successful in most cases, with the index starting at $\gamma < 3$ at low energies (which may well be consistent with an ADAF providing some of the soft excess emission, e.g. \citealt{2012DoneIntrinsicDiscEmission}) and rising to $\gamma > 4$ above 4-6~keV. 

The best-fitting emissivity function varies between sources; we attempt to correlate $\gamma(E)$ with some of the key macro-properties of the AGN (mass, spin, inclination and Eddington ratio), finding either a lack of a correlation or only a tentative hint (with Eddington ratio). In future we will increase the number of sources for correlation studies and extend the energy range by utilising data from {\it NuSTAR}. It will also be extremely interesting to compare these results for a simplified disc-like geometry to a more complex geometry incorporating a more realistic Compton scattering scenario, and to power spectra of XRBs in a variety of spectral states.

\section*{Acknowledgements}

This research was possible through an STFC studentship. We thank Phil Uttley and Peter Boorman for helpful discussion, as well as the anonymous referee for their constructive comments. We also reference the use of python libraries which were essential in this analysis: \texttt{Numpy} \citep{Numpy}, \texttt{Scipy} \citep{Scipy}, \texttt{pandas} \citep{pandas}, \texttt{Matplotlib} \citep{Matplotlib} and \texttt{emcee} \citep{2013emcee}, in addition to \texttt{XSPEC} \citep{1996ArnaudXSPEC}.

\section*{Data Availability}

The data underlying this article are freely accessible in the public HEASARC archives (https://heasarc.gsfc.nasa.gov/).




\bibliographystyle{mnras.bst}
\bibliography{bibliography.bib}  




\appendix


\bsp	
\label{lastpage}
\end{document}